\documentclass[a4paper]{article}

\usepackage{INTERSPEECH2021}
\usepackage{url}
\usepackage[activate]{microtype}
\title{An Improved StarGAN for Emotional Voice Conversion:\\ Enhancing Voice Quality and Data Augmentation}
\name{Xiangheng He$^1$, Junjie Chen$^2$, Georgios Rizos$^1$, Björn W.\ Schuller$^{1,3}$}
\address{
  $^1$GLAM -- Group on Language, Audio, \& Music, Imperial College London, UK\\
  $^2$Department of Computer Science, The University of Tokyo, Japan\\
  $^3$Chair of Embedded Intelligence for Health Care and Wellbeing, University of Augsburg, Germany}
\email{x.he20@imperial.ac.uk}

\begin{document}

\maketitle
\begin{abstract}
Emotional Voice Conversion (EVC) aims to convert the emotional style of a source speech signal to a target style while preserving its content and speaker identity information. Previous emotional conversion studies do not disentangle emotional information from emotion-independent information that should be preserved, thus transforming it all in a monolithic manner and generating audio of low quality, with linguistic distortions. To address this distortion problem, we propose a novel StarGAN framework along with a two-stage training process that separates emotional features from those independent of emotion by using an autoencoder with two encoders as the generator of the Generative Adversarial Network (GAN). The proposed model achieves favourable results in both the objective evaluation and the subjective evaluation in terms of distortion, which reveals that the proposed model can effectively reduce distortion. Furthermore, in data augmentation experiments for end-to-end speech emotion recognition, the proposed StarGAN model achieves an increase of 2\,\% in Micro-F1 and 5\,\% in Macro-F1 compared to the baseline StarGAN model, which indicates that the proposed model is more valuable for data augmentation.
\end{abstract}

\section{Introduction}
Emotional Voice Conversion (EVC) aims to alter the emotional style of a source speech signal to a target style while preserving other emotion-independent information. A major problem of this task is that high-quality parallel data is hard to collect due to substantial labor costs \cite{hantke2017towards}. Recent work proposed to utilise generative adversarial network (GAN) frameworks \cite{goodfellow2014generative, zhu2017unpaired, choi2018stargan} to convert the spectrum features of a source audio signal to the features of the target \cite{luo2019emotional, rizos2020stargan, zhou2020converting, zhou2020seen, shankar2020non}, thus eliminating the need for parallel data. In addition, evidence shows that the GAN-generated audio signals, as a source of augmenting data, are valuable for improving the predictive performance of speech emotion recognition (SER) models \cite{rizos2020stargan,baird2019can}, which not only indicates that the GAN-generated audio signals indeed carry effective emotional information, but also provides a promising data augmentation solution for the SER task \cite{latif2020augmenting,chatziagapi2019data,rizos2019modelling,mertes2020evolutionary}. 

\begin{figure}[t!]
  \centering
  \includegraphics[width=\linewidth]{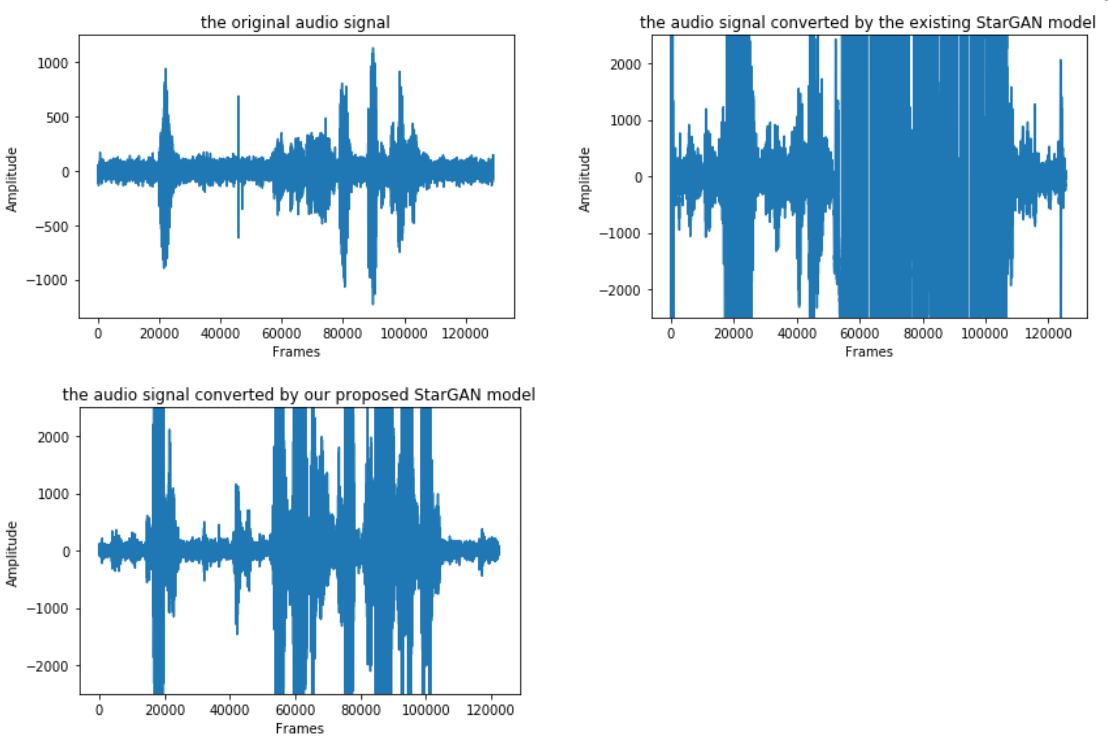}
  \caption{Example waveforms of source and converted audio signals (Ses03F\_script02\_2\_F043.wav in the IEMOCAP dataset). Our proposed StarGAN model introduces fewer artefacts in silent frames compared to the model proposed in \cite{rizos2020stargan}.}
  \label{fig:waveform_example}
\end{figure}

However, comparing the source audio signal waveform with the waveform of the audio signal converted using the existing StarGAN-based EVC model \cite{rizos2020stargan}, we observe that some silent frames in the original waveform were given very high amplitude values in the converted waveform. An example is shown in Figure~\ref{fig:waveform_example}. Since these silent frames represent silence between discrete words or sentences, excessive amplitude values should not be given to these frames in any conversion between emotions. This indicates that the existing StarGAN framework is prone to generating artefacts; we believe that this is because it transforms the input clip in a monolithic manner, instead of focusing on the emotional indices, as desired. This monolithic conversion manner tends to cause the distortion of the linguistic characteristics (e.\,g., cracked voices, or jitter) and damages the overall audio quality. Experimental results of existing EVC models also show that the mean opinion scores (MOS) for audio quality before and after conversion suffer from a significant drop \cite{zhou2020seen}.

We believe that a means to address this problem is to separate emotional features from emotion-independent features. Autoencoder (AE)- or variational autoencoder (VAE)-based feature disentanglement methods provide a possible solution and have been successfully applied to the voice conversion (VC) task \cite{hsu2016voice, liu2018voice, huang2019investigation,qian2019autovc, qian2020f0, huang2020unsupervised}. In \cite{huang2019investigation}, by providing the speaker identity features as a condition to the decoder, the encoder learns to encode only the speaker-independent information in the process of minimising the reconstruction loss. Experimental results from \cite{huang2020unsupervised} show that the degree of disentanglement is positively correlated with the performance of the VC model and can be enhanced by both GANs and the speaker classifier. Inspired by these feature disentanglement methods, we propose to utilise an autoencoder with two encoders as the generator of the StarGAN as well as a two stage training process for the model.

\begin{figure*}[t!]
  \centering
  \includegraphics[width=\linewidth]{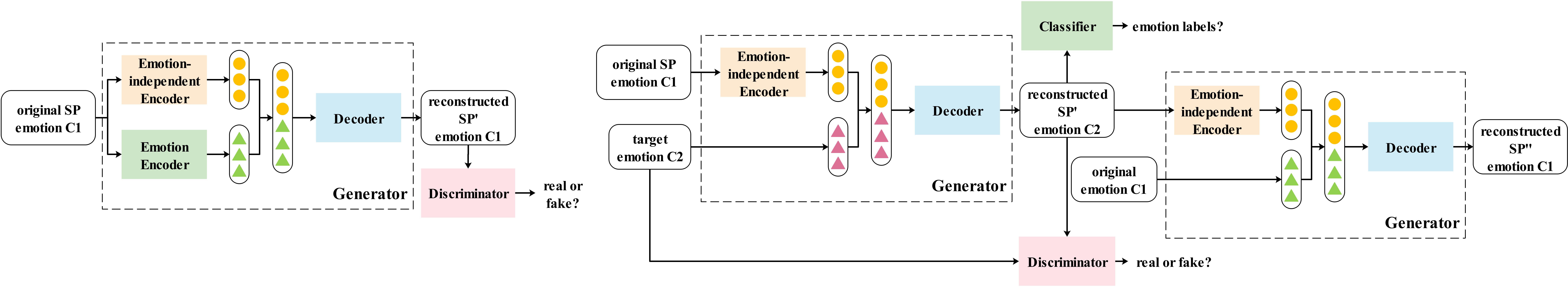}
  \caption{The model structure used in the first (left) and the second (right) stage of training.}
  \label{fig:model_structure}
\end{figure*}

Unlike the existing StarGAN-based EVC model \cite{rizos2020stargan}, we explicitly separate emotion-independent features from emotional features during conversion. When providing the target emotion label to StarGAN, we use continuous emotional features extracted by a pre-trained emotion classifier rather than using one-hot vectors to represent each emotion. Similar continuous emotional features have also been used in \cite{zhou2020seen}. However, their model is based on VAW-GAN in the context of one-to-many emotional conversion, while our model is based on StarGAN and able to complete the many-to-many emotional conversion. Besides, their model simply reconstructs the source spectral envelope without attempting to convert it during the whole training process. The authors only consider the emotion conversion when testing, which causes a mismatch between training and testing. Our model eliminates this mismatch by using a proposed two-stage training process.

The main contributions of this paper are: 1) We propose a novel model structure as well as a two-stage training process based on StarGAN, which explicitly separate emotion-independent features from emotional features during conversion\footnote{Our code is publicly available at: \url{https://github.com/xianghenghe/Improved_StarGAN_Emotional_Voice_Conversion}.}; 2) we perform both the objective evaluation using Mel-cepstral distortion (MCD) \cite{kubichek1993mel} and the subjective evaluation in terms of distortion to ascertain whether this separation of features can indeed reduce the distortion of generated audio signals and improve their overall quality; 3) we replicate the data augmentation experiments of an existing StarGAN-based EVC model \cite{rizos2020stargan} to showcase that our audio quality improvements are not at the expense of emotional conversion performance. 

\section{Improved StarGAN}

Since the harmonic spectral envelope (SP) and the fundamental frequency (F0) contour contain emotion-related cues \cite{10.5555/2553204, yildirim2004acoustic}, we use the WORLD vocoder \cite{morise2016world} to decompose the raw audio signal into SP, F0, and aperiodicity parameters (AP) before converting. We then apply the proposed StarGAN framework to generate the SP with target emotion. We next introduce the two training stages, the model architecture and the final conversion process.

\begin{figure}[t]
  \centering
  \includegraphics[height=70mm,width=64mm]{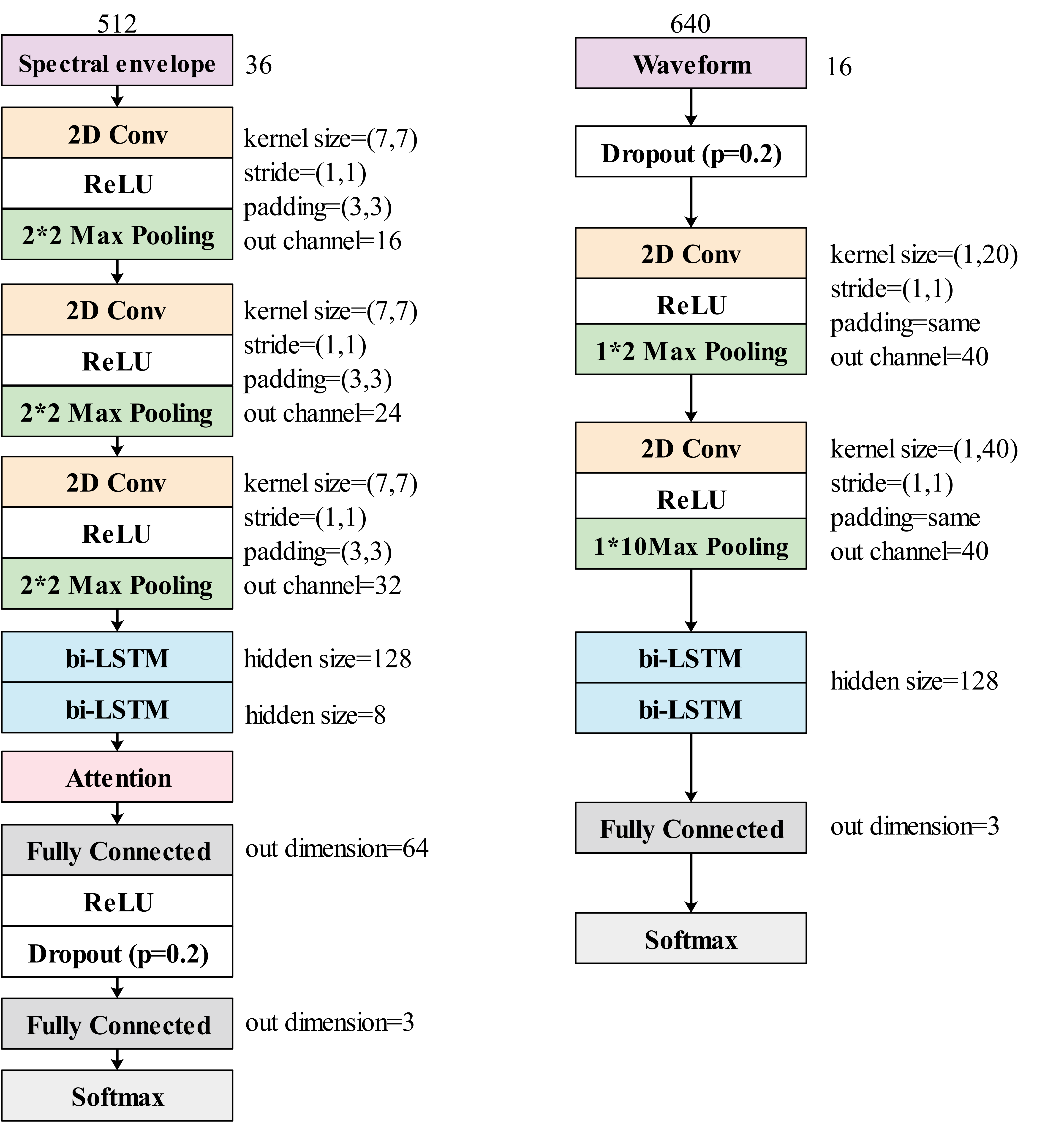}
  \caption{The structure of the pre-trained emotion classifier (left) used as the emotion encoder in the improved StarGAN and the SER model (right) used in the data augmentation experiments.}
  \label{fig:two_classifier}
\end{figure}

\subsection{Training Stage1: Autoencoder training}

In the first training stage, we propose to utilise an autoencoder with two encoders as the GAN generator $\displaystyle G$, which aims to learn an emotion-independent encoder capable of separating emotional features from emotion-independent features. 

As shown in the left part of Figure~\ref{fig:model_structure}, the autoencoder is trained by simply reconstructing the source spectral envelope $\displaystyle SP_{C1}$ with emotion $\displaystyle {C1}$ without attempting to convert it, which means that the target emotion is not provided in this stage. The emotion encoder aims to encode emotional information, where its latent code $\displaystyle E(SP_{C1})$ is provided by a pre-trained emotion classifier (fixed during training stage 1) shown in Figure~\ref{fig:two_classifier}. By providing the emotion features of the source $\displaystyle SP_{C1}$ as a condition, the emotion-independent encoder learns to eliminate emotional information from the input, thus making its latent code $\displaystyle I(SP_{C1})$ emotion independent. The decoder learns to reconstruct $\displaystyle SP_{C1}$ using the concatenated latent codes, which can be formulated as:
\begin{equation}
   SP'_{C1} =G( I( SP_{C1}) ,E( SP_{C1})).
  \label{eq1}
\end{equation}
To make the reconstructed $\displaystyle SP'_{C1}$ more realistic, we train the autoencoder based on adversarial learning. 


\subsection{Training Stage2: StarGAN training}
In the second training stage, the typical StarGAN training is implemented to eliminate the training-testing mismatch in \cite{zhou2020seen} and enable the many-to-many emotion conversion.

As shown in the right part of Figure~\ref{fig:model_structure}, the emotion-independent encoder and the decoder are the generator of the StarGAN, denoted as $\displaystyle G$. We extract the emotion features of audio signals with the same emotion by using the pre-trained emotion classifier and average them as the feature representations of target emotion labels. We denote $\displaystyle \overline{E}_{C2}$ as the representation of target emotion $\displaystyle {C2}$. The generator $\displaystyle G$ first reconstructs the $\displaystyle SP'_{C2}$ from $\displaystyle \overline{E}_{C2}$ and $\displaystyle I(SP_{C1})$ and then reconstructs the $\displaystyle SP''_{C1}$ from $\displaystyle \overline{E}_{C1}$ and $\displaystyle I(SP'_{C2})$:
\begin{equation}
  SP'_{C2} =G(I( SP_{C1}) ,\overline{E}_{C2}),
  \label{eq2}
\end{equation}
\begin{equation}
   SP''_{C1} =G( I( SP'_{C2}) ,\overline{E}_{C1}).
  \label{eq3}
\end{equation}
The generator $\displaystyle G$ tries to minimise the reconstruction loss and fool the discriminator $\displaystyle D$ by generating a more realistic SP. The $\displaystyle D$ tries to maximise the loss between the real SP and the fake SP. The domain classifier $\displaystyle C$ learns to predict the emotion label and helps to enable the many-to-many emotion conversion.


\subsection{Conversion Process at Test Time}
We follow the relative logarithm Gaussian normalised transformation (LGNT) \cite{liu2007high} proposed by \cite{rizos2020stargan} to convert the F0 contour of an audio signal from emotion $\displaystyle {C1}$ to the target emotion $\displaystyle {C2}$. 
\begin{equation}
\label{eq4}
\begin{split}
\log(F0') =((\log(F0) -\mu (F0))\frac{\sigma (F0) +\Delta \sigma _{C_{1} ,\ C_{2}}}{\sigma (F0) \ }\\
+\mu (F0) +\Delta \mu _{C_{1} ,\ C_{2}},
\end{split}
\end{equation}
where $\mu (F0)$ and $\sigma (F0)$ are the mean and variance of $\log(F0)$. $\Delta \mu _{C_{1} ,\ C_{2}}$ and $\Delta \sigma _{C_{1} ,\ C_{2}}$ are the average difference in the mean $\log(F0)$ and the average change in the variance of $\log(F0)$ between emotion $\displaystyle {C1}$ and $\displaystyle {C2}$.

At test time, with the spectral envelope converted by our improved StarGAN model and the F0 contour converted by the relative LGNT, we synthesise the audio signal with the target emotion via the WORLD vocoder.

\subsection{Dataset and Training Details}
We use the Interactive Emotional Dyadic Motion Capture (IEMOCAP) Database \cite{busso2008iemocap} for all our experiments. This dataset has been recorded from ten actors in sessions 1-5, including emotional scripts and improvised hypothetical scenarios. It contains approximately 12 hours of recordings with 9 emotions. The sampling rate of all recordings is 16\,kHz. We utilise three emotions including angry, sad, and happy from both scripted and improvised patterns which contains 2\,435 recordings in totally.

For more details of our improved StarGAN framework, we use the same discriminator and the domain classifier structures as the existing StarGAN-based model \cite{rizos2020stargan, kameoka2018stargan}. The architecture of our emotion-independent encoder and decoder is the same as the generator in \cite{rizos2020stargan, kameoka2018stargan}. The pre-trained classifier used in the first training stage is shown in the left part of 
Figure~\ref{fig:two_classifier}. 
It has the same structure as the domain classifier in the second stage. We extract 36 cepstral coefficients as an approximation to the whole spectral envelope for each training sample. In both the first and the second training stages, we set the batch size to 4 and use the Adam optimiser \cite{kingma2014adam} with the learning rate (for $\displaystyle G$, $\displaystyle D$ and $\displaystyle C$), $\beta1$ and $\beta2$ set to $1\times10^{-4}$, 0.5, and 0.999. We perform 1 generator update after 5 discriminator and domain classifier updates in the StarGAN training stage as in \cite{gulrajani2017improved}. The first training stage lasts for 400 epochs, and we manually stop the second training stage when the loss reaches a plateau (approximately 380 epochs). 

\begin{figure}[t]
  \centering
  \includegraphics[width=\linewidth]{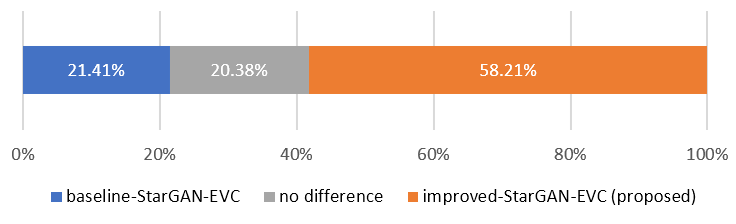}
  \caption{Results of audio preference test.}
  \label{fig:preference}
\end{figure}

\begin{table}[t]
  \caption{Explanation of MOS Scores.}
  \label{tab:MOS_explain}
  \centering
  \begin{tabular}{ccc}
    \toprule
    \textbf{Rating} & \textbf{Quality} & \textbf{Distortion}               \\
    \midrule
    5      & Excellent                 & Imperceptible                \\
    4  & Good   & Just perceptible, but not annoying 
    \\
   3  & Fair   & Perceptible and slightly annoying
    \\
   2  & Poor   & Annoying, but not objectionable
    \\
   1  & Bad   & Very annoying and objectionable
    \\
    \bottomrule
  \end{tabular}
\end{table}

\begin{table}[th]
  \caption{Results of the MOS test with 95\,\% confidence interval.}
  \label{tab:MOS}
  \centering
  \begin{tabular}{ c@{}l  r }
    \toprule
    \multicolumn{2}{c}{\textbf{Models}} & 
 \multicolumn{1}{c}{\textbf{MOS}} \\
    \midrule
    baseline-StarGAN-EVC && $1.952 \pm 0.121$ ~~~  \\
    \textbf{improved-StarGAN-EVC} && $\textbf{2.355} \pm \textbf{0.084}$ ~~~   \\
    \bottomrule
  \end{tabular}
\end{table}

\begin{table*}[t]
  \caption{Results of the MCD for source-target emotion pairs and overall converted audio signals regardless of emotions.}
  \label{tab:MCD}
  \centering
  \begin{tabular}{cccccccc}
    \toprule
    \textbf{MCD(dB)} & \textbf{angry-sad} & \textbf{angry-happy} & \textbf{sad-angry} &  \textbf{sad-happy} &   \textbf{happy-angry} &   \textbf{happy-sad} &  \textbf{overall}     \\
    \midrule
    baseline-StarGAN-EVC  & 4.330 & 4.291 & 6.018 & 5.291 & 5.312 & 4.112 & 4.777
    \\
    \textbf{improved-StarGAN-EVC}  & \textbf{4.265}   & \textbf{4.046}  & \textbf{4.826} & \textbf{3.916} & \textbf{4.719} & \textbf{3.879} & \textbf{4.183}
    \\
    \bottomrule
  \end{tabular}
\end{table*}

\begin{table*}[t]
  \caption{Results of data augmentation experiments. Standard deviations for 10 trials are indicated in parentheses.}
  \label{tab:data_augmentation}
  \centering
  \begin{tabular}{ccc}
    \toprule
    \textbf{Train on} & \textbf{Micro-F1 (\%)} & \textbf{Macro-F1 (\%)}               \\
    \midrule
    sessions1-3      & 59.87 (0.3889)                 & 56.64 (0.3761)                 \\
     sessions1-3+baseline-StarGAN  & 60.37 (0.7502)   & 55.41 (0.7933)
    \\
    \textbf{sessions1-3+improved-StarGAN}  & \textbf{62.62} (1.1956)   & \textbf{60.34} (0.4701)
    \\
    \bottomrule
  \end{tabular}
\end{table*}

\section{Audio Distortion Evaluation}
To ascertain whether the separation of emotional features from emotion-independent features can indeed reduce the distortion of the generated audio signals and improve their overall quality, we perform an objective evaluation using the MCD and a subjective evaluation via the human preference test and the mean opinion scores (MOS) test in terms of distortion.  

\subsection{Subjective Evaluation Experimental Design}
We conduct two listening tests for our human perception experiments, the human preference test and the MOS test in audio distortion. 26 fluent English-speaking subjects participated in all experiments. Each of them was asked to evaluate 60 audio samples (30 pairs), which contains 30 audio samples generated by the three-class StarGAN-based EVC model \cite{rizos2020stargan} (denoted as baseline-StarGAN-EVC) and 30 audio samples generated by our proposed StarGAN-based EVC model (denoted as StarGAN-based EVC). Both models were trained on the IEMOCAP dataset with the same train/validation/test split. Two audio samples of each pair were generated from one source sample in IEMOCAP to a randomly selected target emotion.


\subsection{Subjective Evaluation Results and Discussion}

For the MOS test in terms of distortion, participants were shown a pair of samples for each question and were asked to rate each sample on the MOS scale shown in Table~\ref{tab:MOS_explain}. The MOS test results in terms of distortion with 95\,\% confidence interval are shown in Table~\ref{tab:MOS}. We observe that the improved-StarGAN-EVC model outperforms the baseline-StarGAN-EVC model with a higher MOS score with no confidence interval overlap.

For the audio preference test, participants were asked to compare samples from each pair and choose the one they believe in having less perceived distortion (e.\,g., cracked voices, or jitter). The audio preference test results are summarised in Figure~\ref{fig:preference}. We observe that our proposed model obtained a preference of up to 58.21\,\%, which is almost three times the figure for the baseline-StarGAN-EVC model, indicating that humans clearly favored samples converted by our improved-StarGAN-EVC model. 


\subsection{Objective Evaluation Experimental Design}
We also perform an objective evaluation using MCD, which measures the distortion between two sequences of Mel-cepstral 
coefficients. 
It is a commonly used metric for assessing the quality of generated speech signals in VC \cite{miyoshi2017voice, tian2018average, tobing2019non, tobing2020cyclic} and EVC. We calculate the MCD between the mel cepstra of reference audio signals and that of audio signals converted by both the baseline-StarGAN-EVC and the improved-StarGAN-EVC model for all six source-to-target emotion pairs. 

\subsection{Objective Evaluation Results and Discussion}
Table~\ref{tab:MCD} reports the MCD values of source-target emotion pairs and the overall MCD values for all converted audio signals regardless of emotions. We note that our proposed improved-StarGAN-EVC model outperforms the baseline-StarGAN-EVC model in MCD for all terms, revealing that our proposed model can effectively reduce distortion. Besides, for the sad-happy and sad-angry emotion conversion, our model achieves a distortion reduction of up to 1.375 and 1.192, which indicates that the proposed feature separation method has a much better suppression effect on distortion when the targets are high arousal emotions.

\section{Data Augmentation Evaluation}
Although the audio distortion evaluation experiments show that our proposed feature separation method can effectively reduce distortion, we still need to determine whether or not this improvement in audio quality is at the expense of our model's emotional conversion performance. We perform the data augmentation experiments to validate its emotional conversion performance.

\subsection{Experimental Design}
We use an end-to-end speech emotion recognition (SER) model \cite{zhang2019attention}, which directly takes the waveform as its input to accomplish the data augmentation experiments. The architecture is shown in the right part of Figure~\ref{fig:two_classifier}. We first standardised the raw audio signal by using the mean and standard deviation from the training set. The raw signal was then sampled at an interval of 40\,ms with a step size of 10\,ms. Given the 16\,kHz sampling rate of raw signals, the waveform for each frame is of dimension 640. We 
randomly selected 16 continuous frames for each audio signal and got the fixed-size input waveform of dimension $16\times640$. Note that the SER model here is entirely different from the pre-trained SER model used in our improved StarGAN training. Also, note that the SER model here is slightly different from the model used in \cite{rizos2020stargan} in terms of kernel sizes and filter numbers; that is why the data augmentation results we report for the baseline-StarGAN-EVC are different from the results 
shown in \cite{rizos2020stargan}. Our SER model was trained using the Adam optimiser with the learning rate, $\beta1$ and $\beta2$ set to $1\times10^{-5}$, 0.5, and 0.999. The batch size was set to 4. We employ sessions 1-3 of the IEMOCAP dataset as the original training set, session 4 as the validation set, and session 5 as the testing set. We consider Micro-F1 and Macro-F1 scores to report the performance of the SER model on the testing set. All results are averaged across 10 trials to reduce the effect of initialisation biases. 

\subsection{Results and Discussion}
We report the data augmentation results in Table~\ref{tab:data_augmentation}. We trained the SER model on three datasets, namely the IEMOCAP dataset of sessions 1-3,  IEMOCAP sessions 1-3 augmented with audio signals generated by our proposed improved-StarGAN-EVC model, and IEMOCAP sessions 1-3 augmented with audio signals generated by the baseline-StarGAN-EVC model. We refer to them as sessions1-3, sessions1-3+improved-StarGAN, and sessions1-3+baseline-StarGAN, respectively. We observe that the model trained on sessions1-3+improved-StarGAN achieves higher Micro-F1 and Macro-F1 scores than the model trained only on sessions1-3, which indicates that the audio signals converted by our proposed model indeed carry effective emotional information and are valuable for data augmentation. We also note that the model trained on sessions1-3+improved-StarGAN achieves an absolute increase of 2\,\% in Micro-F1 and 5\,\% in Macro-F1 compared to the model trained on sessions1-3+baseline-StarGAN, which indicates that our model's improvement in audio quality is not at the expense of its emotional conversion performance; on the contrary, the audio signals generated by our proposed model are more consistent with the distribution of the original dataset and are more valuable for data augmentation.

\section{Conclusion}
We proposed a novel StarGAN-based framework and a two-stage training process that explicitly separates the emotional features from emotion-independent features to reduce the audio distortion and improve the emotion conversion performance. Experimental results in the objective evaluation and the subjective evaluation in terms of audio distortion validated that the proposed model can effectively reduce the distortion and improve the voice quality, especially when the targets are high arousal emotions. Results of data augmentation experiments for end-to-end SER indicated that the improvement in audio quality obtained by the proposed model is not at the expense of its emotional conversion performance. The proposed model is more valuable for data augmentation.
Future approaches should also consider verification of pertained speech intelligibility subsequent to conversion, such as by ASR similar to the here shown SER. 


\bibliographystyle{IEEEtran}

\bibliography{mybib}

\begin{thebibliography}{10}
\providecommand{\url}[1]{#1}
\csname url@samestyle\endcsname
\providecommand{\newblock}{\relax}
\providecommand{\bibinfo}[2]{#2}
\providecommand{\BIBentrySTDinterwordspacing}{\spaceskip=0pt\relax}
\providecommand{\BIBentryALTinterwordstretchfactor}{4}
\providecommand{\BIBentryALTinterwordspacing}{\spaceskip=\fontdimen2\font plus
\BIBentryALTinterwordstretchfactor\fontdimen3\font minus
  \fontdimen4\font\relax}
\providecommand{\BIBforeignlanguage}[2]{{%
\expandafter\ifx\csname l@#1\endcsname\relax
\typeout{** WARNING: IEEEtran.bst: No hyphenation pattern has been}%
\typeout{** loaded for the language `#1'. Using the pattern for}%
\typeout{** the default language instead.}%
\else
\language=\csname l@#1\endcsname
\fi
#2}}
\providecommand{\BIBdecl}{\relax}
\BIBdecl

\bibitem{hantke2017towards}
S.~Hantke, Z.~Zhang, and B.~W. Schuller, ``Towards intelligent crowdsourcing
  for audio data annotation: Integrating active learning in the real world.''
  in \emph{INTERSPEECH}, 2017, pp. 3951--3955.

\bibitem{goodfellow2014generative}
I.~J. Goodfellow, J.~Pouget-Abadie, M.~Mirza, B.~Xu, D.~Warde-Farley, S.~Ozair,
  A.~Courville, and Y.~Bengio, ``Generative adversarial networks,'' \emph{arXiv
  preprint arXiv:1406.2661}, 2014.

\bibitem{zhu2017unpaired}
J.-Y. Zhu, T.~Park, P.~Isola, and A.~A. Efros, ``Unpaired image-to-image
  translation using cycle-consistent adversarial networks,'' in
  \emph{Proceedings of the IEEE international conference on computer vision},
  2017, pp. 2223--2232.

\bibitem{choi2018stargan}
Y.~Choi, M.~Choi, M.~Kim, J.-W. Ha, S.~Kim, and J.~Choo, ``Stargan: Unified
  generative adversarial networks for multi-domain image-to-image
  translation,'' in \emph{Proceedings of the IEEE conference on computer vision
  and pattern recognition}, 2018, pp. 8789--8797.

\bibitem{luo2019emotional}
Z.~Luo, J.~Chen, T.~Takiguchi, and Y.~Ariki, ``Emotional voice conversion using
  dual supervised adversarial networks with continuous wavelet transform f0
  features,'' \emph{IEEE/ACM Transactions on Audio, Speech, and Language
  Processing}, vol.~27, no.~10, pp. 1535--1548, 2019.

\bibitem{rizos2020stargan}
G.~Rizos, A.~Baird, M.~Elliott, and B.~Schuller, ``Stargan for emotional speech
  conversion: Validated by data augmentation of end-to-end emotion
  recognition,'' in \emph{ICASSP 2020-2020 IEEE International Conference on
  Acoustics, Speech and Signal Processing (ICASSP)}.\hskip 1em plus 0.5em minus
  0.4em\relax IEEE, 2020, pp. 3502--3506.

\bibitem{zhou2020converting}
K.~Zhou, B.~Sisman, M.~Zhang, and H.~Li, ``Converting anyone's emotion: Towards
  speaker-independent emotional voice conversion,'' \emph{arXiv preprint
  arXiv:2005.07025}, 2020.

\bibitem{zhou2020seen}
K.~Zhou, B.~Sisman, R.~Liu, and H.~Li, ``Seen and unseen emotional style
  transfer for voice conversion with a new emotional speech dataset,''
  \emph{arXiv preprint arXiv:2010.14794}, 2020.

\bibitem{shankar2020non}
R.~Shankar, J.~Sager, and A.~Venkataraman, ``Non-parallel emotion conversion
  using a deep-generative hybrid network and an adversarial pair
  discriminator,'' \emph{arXiv preprint arXiv:2007.12932}, 2020.

\bibitem{baird2019can}
A.~Baird, S.~Amiriparian, and B.~Schuller, ``Can deep generative audio be
  emotional? towards an approach for personalised emotional audio generation,''
  in \emph{2019 IEEE 21st International Workshop on Multimedia Signal
  Processing (MMSP)}.\hskip 1em plus 0.5em minus 0.4em\relax IEEE, 2019, pp.
  1--5.

\bibitem{latif2020augmenting}
S.~Latif, M.~Asim, R.~Rana, S.~Khalifa, R.~Jurdak, and B.~W. Schuller,
  ``Augmenting generative adversarial networks for speech emotion
  recognition,'' \emph{arXiv preprint arXiv:2005.08447}, 2020.

\bibitem{chatziagapi2019data}
A.~Chatziagapi, G.~Paraskevopoulos, D.~Sgouropoulos, G.~Pantazopoulos,
  M.~Nikandrou, T.~Giannakopoulos, A.~Katsamanis, A.~Potamianos, and
  S.~Narayanan, ``Data augmentation using gans for speech emotion
  recognition.'' in \emph{Interspeech}, 2019, pp. 171--175.

\bibitem{rizos2019modelling}
G.~Rizos and B.~Schuller, ``Modelling sample informativeness for deep affective
  computing,'' in \emph{ICASSP 2019-2019 IEEE International Conference on
  Acoustics, Speech and Signal Processing (ICASSP)}.\hskip 1em plus 0.5em minus
  0.4em\relax IEEE, 2019, pp. 3482--3486.

\bibitem{mertes2020evolutionary}
S.~Mertes, A.~Baird, D.~Schiller, B.~W. Schuller, and E.~Andr{\'e}, ``An
  evolutionary-based generative approach for audio data augmentation,'' in
  \emph{2020 IEEE 22nd International Workshop on Multimedia Signal Processing
  (MMSP)}.\hskip 1em plus 0.5em minus 0.4em\relax IEEE, 2020, pp. 1--6.

\bibitem{hsu2016voice}
C.-C. Hsu, H.-T. Hwang, Y.-C. Wu, Y.~Tsao, and H.-M. Wang, ``Voice conversion
  from non-parallel corpora using variational auto-encoder,'' in \emph{2016
  Asia-Pacific Signal and Information Processing Association Annual Summit and
  Conference (APSIPA)}.\hskip 1em plus 0.5em minus 0.4em\relax IEEE, 2016, pp.
  1--6.

\bibitem{liu2018voice}
S.~Liu, J.~Zhong, L.~Sun, X.~Wu, X.~Liu, and H.~Meng, ``Voice conversion across
  arbitrary speakers based on a single target-speaker utterance.'' in
  \emph{Interspeech}, 2018, pp. 496--500.

\bibitem{huang2019investigation}
W.-C. Huang, Y.-C. Wu, C.-C. Lo, P.~L. Tobing, T.~Hayashi, K.~Kobayashi,
  T.~Toda, Y.~Tsao, and H.-M. Wang, ``Investigation of f0 conditioning and
  fully convolutional networks in variational autoencoder based voice
  conversion,'' \emph{arXiv preprint arXiv:1905.00615}, 2019.

\bibitem{qian2019autovc}
K.~Qian, Y.~Zhang, S.~Chang, X.~Yang, and M.~Hasegawa-Johnson, ``Autovc:
  Zero-shot voice style transfer with only autoencoder loss,'' in
  \emph{International Conference on Machine Learning}.\hskip 1em plus 0.5em
  minus 0.4em\relax PMLR, 2019, pp. 5210--5219.

\bibitem{qian2020f0}
K.~Qian, Z.~Jin, M.~Hasegawa-Johnson, and G.~J. Mysore, ``F0-consistent
  many-to-many non-parallel voice conversion via conditional autoencoder,'' in
  \emph{ICASSP 2020-2020 IEEE International Conference on Acoustics, Speech and
  Signal Processing (ICASSP)}.\hskip 1em plus 0.5em minus 0.4em\relax IEEE,
  2020, pp. 6284--6288.

\bibitem{huang2020unsupervised}
W.-C. Huang, H.~Luo, H.-T. Hwang, C.-C. Lo, Y.-H. Peng, Y.~Tsao, and H.-M.
  Wang, ``Unsupervised representation disentanglement using cross domain
  features and adversarial learning in variational autoencoder based voice
  conversion,'' \emph{IEEE Transactions on Emerging Topics in Computational
  Intelligence}, vol.~4, no.~4, pp. 468--479, 2020.

\bibitem{kubichek1993mel}
R.~Kubichek, ``Mel-cepstral distance measure for objective speech quality
  assessment,'' in \emph{Proceedings of IEEE Pacific Rim Conference on
  Communications Computers and Signal Processing}, vol.~1.\hskip 1em plus 0.5em
  minus 0.4em\relax IEEE, 1993, pp. 125--128.

\bibitem{10.5555/2553204}
B.~Schuller and A.~Batliner, \emph{Computational Paralinguistics: Emotion,
  Affect and Personality in Speech and Language Processing}.\hskip 1em plus
  0.5em minus 0.4em\relax John Wiley \& Sons, 2013.

\bibitem{yildirim2004acoustic}
S.~Yildirim, M.~Bulut, C.~M. Lee, A.~Kazemzadeh, Z.~Deng, S.~Lee, S.~Narayanan,
  and C.~Busso, ``An acoustic study of emotions expressed in speech,'' in
  \emph{Eighth International Conference on Spoken Language Processing}, 2004.

\bibitem{morise2016world}
M.~Morise, F.~Yokomori, and K.~Ozawa, ``World: a vocoder-based high-quality
  speech synthesis system for real-time applications,'' \emph{IEICE
  TRANSACTIONS on Information and Systems}, vol.~99, no.~7, pp. 1877--1884,
  2016.

\bibitem{liu2007high}
K.~Liu, J.~Zhang, and Y.~Yan, ``High quality voice conversion through
  phoneme-based linear mapping functions with straight for mandarin,'' in
  \emph{Fourth International Conference on Fuzzy Systems and Knowledge
  Discovery (FSKD 2007)}, vol.~4.\hskip 1em plus 0.5em minus 0.4em\relax IEEE,
  2007, pp. 410--414.

\bibitem{busso2008iemocap}
C.~Busso, M.~Bulut, C.-C. Lee, A.~Kazemzadeh, E.~Mower, S.~Kim, J.~N. Chang,
  S.~Lee, and S.~S. Narayanan, ``Iemocap: Interactive emotional dyadic motion
  capture database,'' \emph{Language resources and evaluation}, vol.~42, no.~4,
  pp. 335--359, 2008.

\bibitem{kameoka2018stargan}
H.~Kameoka, T.~Kaneko, K.~Tanaka, and N.~Hojo, ``Stargan-vc: Non-parallel
  many-to-many voice conversion using star generative adversarial networks,''
  in \emph{2018 IEEE Spoken Language Technology Workshop (SLT)}.\hskip 1em plus
  0.5em minus 0.4em\relax IEEE, 2018, pp. 266--273.

\bibitem{kingma2014adam}
D.~P. Kingma and J.~Ba, ``Adam: A method for stochastic optimization,''
  \emph{arXiv preprint arXiv:1412.6980}, 2014.

\bibitem{gulrajani2017improved}
I.~Gulrajani, F.~Ahmed, M.~Arjovsky, V.~Dumoulin, and A.~Courville, ``Improved
  training of wasserstein gans,'' \emph{arXiv preprint arXiv:1704.00028}, 2017.

\bibitem{miyoshi2017voice}
H.~Miyoshi, Y.~Saito, S.~Takamichi, and H.~Saruwatari, ``Voice conversion using
  sequence-to-sequence learning of context posterior probabilities,''
  \emph{arXiv preprint arXiv:1704.02360}, 2017.

\bibitem{tian2018average}
X.~Tian, J.~Wang, H.~Xu, E.~S. Chng, and H.~Li, ``Average modeling approach to
  voice conversion with non-parallel data.'' in \emph{Odyssey}, vol. 2018,
  2018, pp. 227--232.

\bibitem{tobing2019non}
P.~L. Tobing, Y.-C. Wu, T.~Hayashi, K.~Kobayashi, and T.~Toda, ``Non-parallel
  voice conversion with cyclic variational autoencoder,'' \emph{arXiv preprint
  arXiv:1907.10185}, 2019.

\bibitem{tobing2020cyclic}
P.~L. Tobing, T.~Hayashi, Y.-C. Wu, K.~Kobayashi, and T.~Toda, ``Cyclic
  spectral modeling for unsupervised unit discovery into voice conversion with
  excitation and waveform modelingh,'' in \emph{Proceedings of INTERSPEECH},
  vol. 2020, 2020.

\bibitem{zhang2019attention}
Z.~Zhang, B.~Wu, and B.~Schuller, ``Attention-augmented end-to-end multi-task
  learning for emotion prediction from speech,'' in \emph{ICASSP 2019-2019 IEEE
  International Conference on Acoustics, Speech and Signal Processing
  (ICASSP)}.\hskip 1em plus 0.5em minus 0.4em\relax IEEE, 2019, pp. 6705--6709.

\end{thebibliography}

\end{document}